\documentclass[11 pt]{article}%
\usepackage{amssymb}
\usepackage{graphicx}
\usepackage{amsmath}
\usepackage{amscd}
\usepackage{amsfonts}
\usepackage{cite}%
\providecommand{\U}[1]{\protect\rule{.1in}{.1in}}
 \evensidemargin0in \oddsidemargin0in \topmargin10pt
\begin{document}

\title{Photon-subtracted squeezed thermal state: nonclassicality and
decoherence\thanks{{\small Work supported by the National Natural Science
Foundation of China under grant 10775097 and 10874174, and the Research
Foundation of the Education Department of Jiangxi Province of China (No.
GJJ10097).}}}
\author{Li-yun Hu$^{1,2}$\thanks{{\small Corresponding author. E-mail:
hlyun2008@126.com.}}, Xue-xiang Xu$^{1,2}$, Zi-sheng Wang$^{1,2}$ and Xue-fen
Xu$^{3}$\\$^{1}${\small College of Physics \& Communication Electronics, Jiangxi Normal
University, Nanchang 330022, China}\\$^{2}${\small Key Laboratory of Optoelectronic and Telecommunication of
Jiangxi, Nanchang, Jiangxi 330022, China}\\$^{3}${\small School of Mathematics and Physics, Jiangsu Teachers University
of Technology, Changzhou, Jiangsu 213001, China}}
\maketitle

\begin{abstract}
{\small We investigate nonclassical properties of the field states generated
by subtracting any number photon from the squeezed thermal state (STS). It is
found that the normalization factor of photon-subtracted STS (PSSTS) is a
Legendre polynomial of squeezing parameter }${\small r}${\small \ and average
photon number }$\bar{n}$ {\small of thermal state. Expressions of several
quasi-probability distributions of PSSTS are derived analytically.
Furthermore, the nonclassicality is discussed in terms of the negativity of
Wigner function (WF). It is shown that the WF of single PSSTS always has
negative values if }$\bar{n}<\sinh^{2}r${\small \ at the phase space center.
The decoherence effect on PSSTS is then included by analytically deriving the
time evolution of WF. The results show that the WF of single PSSTS has
negative value if }$2\kappa t<\ln\{1-(2\bar{n}+1)(\bar{n}-\sinh^{2}%
r)${\small /}$[(2\mathfrak{N}+1)(\bar{n}\cosh2r+\sinh^{2}r)]\}${\small , which
is dependent not only on average number }$\mathfrak{N}${\small \ of
environment, but also on }$\bar{n}$ {\small and }$r${\small . }

\end{abstract}

Keywords: Nonclassicality, decoherence, Photon-subtraction, Squeezed thermal state

PACS number(s): 42.50.Dv, 03.65.Wj, 03.67.Mn

\section{Introduction}

Nonclassical Gaussian states play an important role in quantum information
processing with continuous variables, such as teleportation, dense coding, and
quantum cloning. In a quantum optics laboratory, Gaussian states have been
generated but there is some\ limitation in using them for various tasks of
quantum information procession \cite{r1}. For example, when a two-mode
squeezed vacuum state (a Gaussian state) with low squeezing is used as an
entangled resource to realize quantum teleportation, the average fidelity is
just more $\left(  8\pm2\right)  \%$ than the classical limits. On the other
hand, it is possible to generate and manipulate various nonclassical optical
fields by subtracting or adding photons from/to traditional quantum states or
Gaussian states, which are useful ways to conditionally manipulate
nonclassical state of optical field \cite{2a,2b,2c,2d,2e,2f,2g,2h,2i,2j}.
Recently, subtracting or adding photon states have received more attention
from both experimentalists and theoreticians \cite{r6,r11,r12,r13,r13a,r13b}.
One of reasons is that photon subtraction can be applied to improve
entanglement between Gaussian states \cite{r7,r14}, loophole-free tests of
Bell's inequality \cite{r15,r16}, and quantum computing \cite{r17}. Thus the
photon subtraction (a non-Gaussian operation) can satisfy the requirement of
quantum information protocols for long-distance communication. Nevertheless,
the photon addition and subtraction have been successfully demonstrated
experimentally for probing quantum commutation rules by Parigi \textit{et al}.
\cite{8,8a}. In fact, they have implemented simple alternated sequences of
photon creation (addition) and annihilation (subtraction) on a thermal field
and observed the noncommutativity of the creation and annihilation operators.

In addition, Olivares et al. \cite{r17a} theoretically discussed the relation
between the photon subtracted squeezed vacuum (PSSV), as an output state
passing through a beamsplitter, and two parameters (the transmissivity of
beamsplitter and the photodetection quantum efficiency). Then the case of
two-mode photon-subtraction is also further discussed in the presence of noise
\cite{r17b,r17c}. Kitagawa et al \cite{Ki} investigated the degree of
entanglement for non-Gaussian mixed (pure) states generated by photon
subtraction from two-mode squeezed vacuum states with on-off photon detectors.
For the single PSSV, furthermore, its nonclassical properties and decoherence
was investigated theoretically in two different decoherent channels (amplitude
decay and phase damping) by Biswas and Agarwal \cite{2d}. They indicated that
the WF losses its non-Gaussian nature and becomes Gaussian at long times in
amplitude decay case. Recently, it is found that consecutive applications of
photon subtraction (or subtracting a well-defined number of photons) from a
squeezed vacuum state result in the generation of a squeezed superpositions of
coherent state (SSCS) with nearly the perfect fidelity regardless of the
number of photons subtracted \cite{2e}. The amplitude of the SSCS increases as
the number of the subtracted photons gets larger.

It is interesting in noticing that single-mode displaced-squeezed thermal
state can be considered as a generalized Gaussian state, which has received
more attention \cite{r2,17,18,19,20}. For example, phase estimations for
squeezed thermal states (STSs) and displaced thermal states are presented
\cite{r2}, which shows that a larger temperature can enhance the estimation
fidelity for the former. Another example is, for Gaussian squeezed states of
light, that a scheme is also presented experimentally to measure its
squeezing, purity and entanglement \cite{19,20}. To our knowledge, however,
the investigation of photon subtraction from STS (even for single photon
subtraction case) has not been previously addressed (especially when this
state interacts with its surrounding environment). In addition, the exact
threshold value of the decay time has not been explicity given.\textbf{ }In
this paper, we focus on any number photon-subtracted single-mode STS (PSSTS),
which is optically produced single-mode non-Gaussian states, and explore
theoretically its nonclassical properties and decoherence in a thermal channel
by deriving analytically some expressions, such as normalized constant,
photon-number distribution and Wigner function (WF). For single PSSTS, it is
shown that the WF of single PSSTS always has the negative values under the
condition of $\bar{n}<\sinh^{2}r$\ at the phase space center ($\bar{n}$ and
$r$ are an average number of thermal state and a squeezing parameter,
respectively),\ and that the threshold value of the decay time is dependent
not only on the average number of environment, but also on $\bar{n}$ and $r$.

In section II, we introduce the single-mode PSSTS, where the normalized factor
turns out to be a Legendre polynomial with a remarkable result. In Sec. III,
the nonclassical properties of the PSSTS, such as Mandel's $Q$-parameter, and
distribution of photon number (related to a Legendre polynomial), are
calculated analytically and then be discussed in details. In Sec. IV, the
explicitly analytical expressions of quasiprobability distributions for PSSTS,
such as P-distribution, Q-function and WF of the PSSTS, are derived by using
the Weyl ordered operators' invariance under a similar transformations. Then
we derive an explicitly analytical expression of time evolution of WF for the
arbitrary PSSTS in the thermal channel and discuss the loss of nonclassicality
in reference of the negativity of WF in Sec. V. It is found that the threshold
value of decay time corresponding to the transition of WF from partial
negative to completely positive definite is obtained at the center of the
phase space, which is not only dependent on the average number $\mathfrak{N}$
of environment, but also on $\bar{n}$ and $r$. We show that the WF for single
PSSTS has always negative value if the decay time $\kappa t<\frac{1}{2}%
\ln\{1-(2\bar{n}+1)(\bar{n}-\sinh^{2}r)/[(2\mathfrak{N}+1)(\bar{n}%
\cosh2r+\sinh^{2}r)]\}$ (see Eq.(\ref{f42}) below), where $\kappa$ denotes a
dissipative coefficient of interacting with the environment. Sec. VI is
devoted to calculating the fidelity between the PSSTS and the STS. It is shown
that the fidelity decreases monotonously with the increment of both
photon-subtraction number $m$ and the squeezing parameter $r$. We end with the
main conclusions of our work.

\section{Photon-subtraction squeezed thermal state}

At first, let's introduce the photon-subtraction squeezed thermal state
(PSSTS). For a squeezed thermal field, its density operator is
\begin{equation}
\rho_{s}=S(r)\rho_{c}S^{\dagger}(r), \label{f1}%
\end{equation}
where $S(r)=\exp[r(a^{\dagger2}-a^{2})/2]=\exp[-$i$r(QP+PQ)/2]$ is the
squeezing operator \cite{23,24} with squeezing parameter $r$, here the
coordinate operators $Q=(a+a^{\dagger})/\sqrt{2}$ and the momentum operators
$P=(a-a^{\dagger})/(\sqrt{2}\mathtt{i)}$ $(\left[  a,a^{\dagger}\right]  =1)$
are introduced as functions of create and annihilation operators $a^{\dagger}$
and $a$, respectively, and $\rho_{c}$ is a density operator of thermal state,
\begin{equation}
\rho_{c}=(1-e^{\sigma})e^{\sigma a^{\dagger}a},\text{ }\sigma=-\frac
{\hbar\omega}{kT}, \label{f2}%
\end{equation}
where $k$ is a Boltzmann constant, and the temperature $T$ is qualified to be
a density operator of thermal (chaotic) field with tr$\rho_{c}=1$. \ Using the
operator identity \cite{28,29}
\begin{equation}
e^{\sigma a^{\dagger}a}=\colon\exp[(e^{\sigma}-1)a^{\dagger}a]\colon=\frac
{2}{e^{\sigma}+1}%
%TCIMACRO{\QATOP{:}{:}}%
%BeginExpansion
\genfrac{}{}{0pt}{}{:}{:}%
%EndExpansion
\exp\left\{  \frac{e^{\sigma}-1}{e^{\sigma}+1}\left(  Q^{2}+P^{2}\right)
\right\}
%TCIMACRO{\QATOP{\colon}{\colon}}%
%BeginExpansion
\genfrac{}{}{0pt}{}{\colon}{\colon}%
%EndExpansion
, \label{f3}%
\end{equation}
where these two symbols $\colon$ $\colon$ and $%
%TCIMACRO{\QATOP{:}{:}}%
%BeginExpansion
\genfrac{}{}{0pt}{}{:}{:}%
%EndExpansion%
%TCIMACRO{\QATOP{:}{:}}%
%BeginExpansion
\genfrac{}{}{0pt}{}{:}{:}%
%EndExpansion
$denote normal ordering and Weyl ordering, respectively, and using the Weyl
ordering invariance under similarity transformations \cite{28,29}, which means
that%
\begin{equation}
S%
%TCIMACRO{\QATOP{:}{:}}%
%BeginExpansion
\genfrac{}{}{0pt}{}{:}{:}%
%EndExpansion
\left(  \circ\circ\circ\right)
%TCIMACRO{\QATOP{:}{:}}%
%BeginExpansion
\genfrac{}{}{0pt}{}{:}{:}%
%EndExpansion
S^{-1}=%
%TCIMACRO{\QATOP{:}{:}}%
%BeginExpansion
\genfrac{}{}{0pt}{}{:}{:}%
%EndExpansion
S\left(  \circ\circ\circ\right)  S^{-1}%
%TCIMACRO{\QATOP{:}{:}}%
%BeginExpansion
\genfrac{}{}{0pt}{}{:}{:}%
%EndExpansion
, \label{f4}%
\end{equation}
as if the \textquotedblleft fence" $%
%TCIMACRO{\QATOP{:}{:}}%
%BeginExpansion
\genfrac{}{}{0pt}{}{:}{:}%
%EndExpansion%
%TCIMACRO{\QATOP{:}{:}}%
%BeginExpansion
\genfrac{}{}{0pt}{}{:}{:}%
%EndExpansion
$did not exist, so $S$ can pass through it, as well as the technique of
integration within an ordered product of operators (IWOP), one can convert
$\rho_{s}$ to its normally ordered Gaussian form \cite{29} (see Appendix A),
i.e.,
\begin{equation}
\rho_{s}=\frac{1}{\tau_{1}\tau_{2}}\colon\exp\left\{  -\frac{Q^{2}}{2\tau
_{1}^{2}}-\frac{P^{2}}{2\tau_{2}^{2}}\right\}  \colon, \label{f5}%
\end{equation}
where$\allowbreak$
\begin{equation}
2\tau_{1}^{2}=(2\bar{n}+1)e^{2r}+1,2\tau_{2}^{2}=(2\bar{n}+1)e^{-2r}+1,
\label{f6}%
\end{equation}
which leads to the following relations,
\begin{align}
\tau_{1}^{2}-\tau_{2}^{2}  &  =(2\bar{n}+1)\sinh2r,\label{f6a}\\
\tau_{1}^{2}+\tau_{2}^{2}  &  =(2\bar{n}+1)\cosh2r+1,\label{f6b}\\
\tau_{1}^{2}\tau_{2}^{2}  &  =\bar{n}^{2}+\left(  2\bar{n}+1\right)
\allowbreak\cosh^{2}r, \label{f6c}%
\end{align}
and $\bar{n}=\mathtt{tr}\left(  \rho_{c}a^{\dagger}a\right)  =(e^{-\sigma
}-1)^{-1}$\cite{30} denotes the average photon number of thermal (chaotic)
field $\rho_{c}$ in Eq. (\ref{f2}). The form in Eq.(\ref{f5}) is similar to
the bivariate normal distribution in statistics, which is useful for us to
further derive the marginal distributions of $\rho_{s}$.

Theoretically, the PSSTS can be obtained by repeatedly operating the photon
annihilation operator $a$\ on a squeezed thermal state, so its density
operator is given by%
\begin{equation}
\rho=C_{m}^{-1}a^{m}\rho_{s}a^{\dag m}, \label{f7}%
\end{equation}
where $m$ is the subtracted photon number (a non-negative integer), and
$C_{m}$ is a normalized constant with (see Appendix B)
\begin{equation}
C_{m}=\mathtt{Tr}(a^{m}\rho_{s}a^{\dag m})=m!D^{m/2}P_{m}\left(  B/\sqrt
{D}\right)  , \label{f8}%
\end{equation}
which indicates that $C_{m}$ is just related to Legendre polynomial
$P_{m}\left(  x\right)  $ (see Appendix B (B10)), and
\begin{align}
B  &  =\frac{1}{2}\left[  \left(  2\bar{n}+1\right)  \cosh2r-1\right]
,\label{f10}\\
D  &  =\bar{n}^{2}-\left(  2\bar{n}+1\right)  \sinh^{2}r. \label{f9}%
\end{align}
It is noted that, for the case of no-photon-subtraction with $m=0$, $C_{0}=1$
as expected. Under the case of $m$-photon-subtraction thermal state with
$B=\bar{n}$, $D=\bar{n}^{2},$ and $P_{m}\left(  1\right)  =1$, $C_{m}%
=m!\bar{n}^{m}.$ The same result as Eq.(24) can be found in Ref.\cite{31}.

Here we should point out that, as Agarwal et al introduced the excitations on
a coherent state by repeated application of the photon creation operator on
the coherent state \cite{31b}, we introduce theoretically the PSSTS
(\ref{f7}). In realistic situations, one the other hand, the photon
subtraction would be done by on/off detector and the tapping beam splitters
with a non-unity transmittance, which leads to a generated mixed state. For
various schemes for generating photon subtraction, one can refer to
Refs.\cite{r1,Ki,31c}.

\section{Nonclassical properties of PSSTS}

\subsection{Mandel's $Q$-parameter}

The analytical expression of $C_{m}$ is of importance for further
investigating the properties of PSSTS. For instance, one can easily calculate
\begin{align}
\left\langle a^{\dag}a\right\rangle  &  =\mathtt{Tr}(C_{m}^{-1}a^{m+1}\rho
_{s}a^{\dag m+1})=\frac{C_{m+1}}{C_{m}},\label{f11}\\
\left\langle a^{\dag2}a^{2}\right\rangle  &  =\mathtt{Tr}(C_{m}^{-1}%
a^{m+2}\rho_{s}a^{\dag m+2})=\frac{C_{m+2}}{C_{m}}, \label{f12}%
\end{align}
thus the Mandel's $Q$-parameter is given by%
\begin{equation}
Q_{M}=\frac{\left\langle a^{\dag2}a^{2}\right\rangle }{\left\langle a^{\dag
}a\right\rangle }-\left\langle a^{\dag}a\right\rangle =\frac{C_{m+2}}{C_{m+1}%
}-\frac{C_{m+1}}{C_{m}}, \label{f13}%
\end{equation}
which measures the deviation of the variance of the photon number distribution
of the field state under consideration from the Poissonian distribution of the
coherent state. If $Q_{M}=0$ we say the field has Poissonian photon
statistics, while for $Q_{M}>0$ ($Q_{M}<0$), the field has super-(sub-)
Poissonian photon statistics. It is well-known that the negativity of the
$Q_{M}$-parameter refers to sub-Poissonian statistics of the state. But a
state may be nonclassical even though $Q_{M}$ is positive\textbf{ }as pointed
out in Ref.\cite{31a}. This case is true for the present state.\textbf{ }In
fact, if $Q_{M}$ is positive, it does not mean that the state is classical. In
such cases, we have to use other parameters to test the
non-classicality\textbf{ }\cite{31a}\textbf{.} From Fig.1, one can see clearly
that for odd number $m$, $Q_{M}$\ becomes negative when the squeezing
parameter $r$ is less than a certain threshold value which decreases as $m$
increases. Differently from the case of odd number $m$, $Q_{M}$ is always
positive for even number $m$. It is necessary to emphasize that the Wigner
function (WF) has negative region for all $r,$ and thus the PSSTS is
nonclassical. In addition, when the average photon number $\bar{n}$ is larger
than a certain threshold value, $Q_{M}$ is also always positive. Without loss
of generality, thus, we consider only the (ideal) PSSTS in a thermal channel
in our following work.

\begin{figure}[ptb]
\label{Fig1}
\centering\includegraphics[width=10cm]{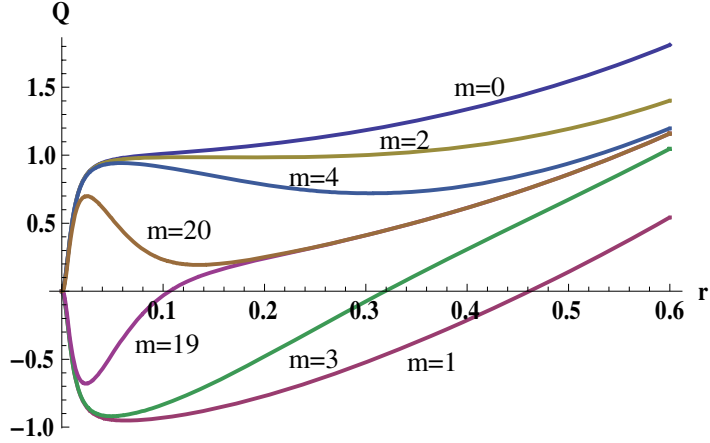}\caption{{\protect\small (Color
online) The }$Q${\protect\small -parameter as the function of squeezing
parameter }$r${\protect\small \ for different }$m=0,1,2,3,4,19,20.$}%
\end{figure}

\subsection{Photon-number distribution of PSSTS}

Next we discuss the photon-number distribution (PND) of PSSTS. Noticing
$a^{\dag m}\left\vert n\right\rangle =\sqrt{(m+n)!/n!}\left\vert
m+n\right\rangle $ and using the un-normalized coherent state $\left\vert
\alpha\right\rangle =\exp[\alpha a^{\dag}]\left\vert 0\right\rangle
$,\cite{32,33} leading to $\left\vert n\right\rangle =\frac{1}{\sqrt{n!}}%
\frac{\mathtt{d}^{n}}{\mathtt{d}\alpha^{n}}\left\vert \alpha\right\rangle
\left\vert _{\alpha=0}\right.  ,$ $\left(  \left\langle \beta\right.
\left\vert \alpha\right\rangle =e^{\alpha\beta^{\ast}}\right)  $, as well as
the normal ordering form of $\rho_{s}$ in Eq. (\ref{f5}), the probability of
finding $n$ photons in the field is given by
\begin{align}
\mathcal{P}(n)  &  =\left\langle n\right\vert \rho\left\vert n\right\rangle
=C_{m}^{-1}\left\langle n\right\vert a^{m}\rho_{s}a^{\dag m}\left\vert
n\right\rangle \nonumber\\
&  =\frac{(m+n)!}{n!C_{m}}\left\langle m+n\right\vert \rho_{s}\left\vert
m+n\right\rangle \nonumber\\
&  =\frac{C_{m}^{-1}}{n!\tau_{1}\tau_{2}}\frac{d^{m+n}}{d\beta^{\ast m+n}%
}\frac{d^{m+n}}{d\alpha^{m+n}}\left.  \left\langle \beta\right\vert \colon
\exp\left\{  -\frac{\left(  a+a^{\dag}\right)  ^{2}}{4\tau_{1}^{2}}%
+\frac{\left(  a-a^{\dag}\right)  ^{2}}{4\tau_{2}^{2}}\right\}  \colon
\left\vert \alpha\right\rangle \right\vert _{\alpha=\beta^{\ast}=0}\nonumber\\
&  =\frac{C_{m}^{-1}}{n!\tau_{1}\tau_{2}}\frac{d^{m+n}}{d\beta^{\ast m+n}%
}\frac{d^{m+n}}{d\alpha^{m+n}}\left.  \exp\left\{  -\frac{\left(  \alpha
+\beta^{\ast}\right)  ^{2}}{4\tau_{1}^{2}}+\frac{\left(  \alpha-\beta^{\ast
}\right)  ^{2}}{4\tau_{2}^{2}}+\alpha\beta^{\ast}\right\}  \right\vert
_{\alpha=\beta^{\ast}=0}\nonumber\\
&  =\frac{C_{m}^{-1}}{n!\tau_{1}\tau_{2}}\frac{d^{2m+2n}}{d\beta^{\ast
m+n}d\alpha^{m+n}}\left.  \exp\left\{  A_{1}\alpha\beta^{\ast}+A_{2}\left(
\alpha^{2}+\beta^{\ast2}\right)  \right\}  \right\vert _{\alpha=\beta^{\ast
}=0}, \label{f14}%
\end{align}
where $A_{1}$ and $A_{2}$ are defined by%
\begin{align}
A_{1}  &  =1-\frac{1}{2\tau_{2}^{2}}-\frac{1}{2\tau_{1}^{2}}=\frac
{\allowbreak\bar{n}\left(  \bar{n}+1\right)  }{\bar{n}^{2}+\left(  2\bar
{n}+1\right)  \allowbreak\cosh^{2}r},\text{ }\\
A_{2}  &  =\frac{1}{4\tau_{2}^{2}}-\frac{1}{4\tau_{1}^{2}}=\frac{(2\bar
{n}+1)\sinh2r}{4\left(  \bar{n}^{2}+\left(  2\bar{n}+1\right)  \allowbreak
\cosh^{2}r\right)  }.
\end{align}
In a similar way to deriving Eq.(B11), we finally obtain%
\begin{align}
\mathcal{P}(n)  &  =\frac{C_{m}^{-1}}{n!\tau_{1}\tau_{2}}\frac{d^{2m+2n}%
}{d\beta^{\ast m+n}d\alpha^{m+n}}\left.  \exp\left\{  -\left(  i\sqrt{A_{2}%
}\alpha\right)  ^{2}-\left(  -i\sqrt{A_{2}}\beta^{\ast}\right)  ^{2}%
+\frac{A_{1}}{A_{2}}\left(  i\sqrt{A_{2}}\alpha\right)  \left(  -i\sqrt{A_{2}%
}\beta^{\ast}\right)  \right\}  \right\vert _{\alpha=\beta^{\ast}%
=0}\nonumber\\
&  =\frac{\left(  A_{2}\right)  ^{m+n}C_{m}^{-1}}{n!\tau_{1}\tau_{2}}%
\frac{d^{2m+2n}}{d\beta^{\ast m+n}d\alpha^{m+n}}\left.  \exp\left\{
-\alpha^{2}-\beta^{\ast2}+\frac{A_{1}}{A_{2}}\alpha\beta^{\ast}\right\}
\right\vert _{\alpha=\beta^{\ast}=0}\nonumber\\
&  =\frac{\left(  m+n\right)  !}{n!\tau_{1}\tau_{2}C_{m}}E^{(m+n)/2}%
P_{m+n}\left(  A_{1}/\sqrt{E}\right)  , \label{f15}%
\end{align}
where $P_{m+n}\left(  x\right)  $ is Legendre polynomial in (B10), and%

\begin{equation}
E=A_{1}^{2}-4A_{2}^{2}=\frac{\bar{n}^{2}-\left(  2\bar{n}+1\right)  \sinh
^{2}r}{\bar{n}^{2}+\left(  2\bar{n}+1\right)  \allowbreak\cosh^{2}r}.
\label{f15a}%
\end{equation}
In particular, when $m=0$ $(C_{0}=1),$ Eq.(\ref{f15}) reduces to
\begin{equation}
\mathcal{P}(n)=\frac{E^{n/2}}{\tau_{1}\tau_{2}}P_{n}\left(  A_{1}/\sqrt
{E}\right)  , \label{f16}%
\end{equation}
which is just the PND of STS which seems a new result; while for $r=0$
($\tau_{1}\tau_{2}=\bar{n}+1,A_{1}=\bar{n}/(\bar{n}+1)$, $E=\bar{n}^{2}%
/(\bar{n}+1)^{2}$,\ $P_{m}\left(  1\right)  =1$, $C_{m}=m!\bar{n}^{m},$),
Eq.(\ref{f15}) becomes%
\begin{equation}
\mathcal{P}(n)=\frac{\left(  m+n\right)  !}{m!n!\bar{n}^{m}}\frac{\bar
{n}^{m+n}}{\left(  \bar{n}+1\right)  ^{m+n+1}}, \label{f17}%
\end{equation}
which is the PND of $m-$photon-subtracted thermal state which also seems a new
result, and the PND ($\bar{n}^{n}/\left(  \bar{n}+1\right)  ^{n+1}$) of
thermal state without photon-subtraction \cite{34,35}.

In Fig.2, the PND is shown for different values ($\bar{n};r$) and $m$, from
which we can see that by subtracting photons, we have been able to move the
peak from zero photons to nonzero photons (see Fig.2 (a)-(c)). The position of
peak depends on how many photons are annihilated and how much the state is
squeezed initially. In addition, the PND mainly shifts to the bigger number
states and becomes more \textquotedblleft flat" and \textquotedblleft wide"
with the increasing parameter $r$ and the average photon-number of thermal
field $\rho_{c}$ (see Fig.2 (b) and (d)).

\begin{figure}[ptb]
\label{Fig2}
\centering\includegraphics[width=12cm]{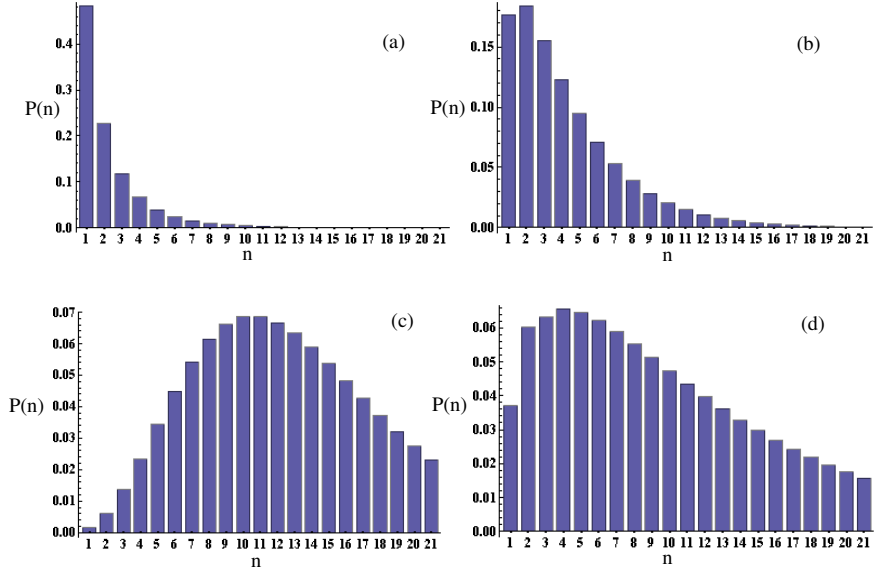}\caption{{\protect\small (Color
online) Photon-number distributions of PASSTS with \={n}=1 for (a) r=0.3,m=0;
(b) r=0.3,m=1; (c) r=0.3,m=5; and (d) r=0.8;m=1.}}%
\end{figure}

\section{Quasiprobability distributions of PSSTS}

In this section, several quasiprobability distributions of PSSTS are derived
in order to provide a convenient way for studying the nonclassical properties
of fields.

\subsection{P distribution}

We first calculate the Glauber-Sudarshan P distribution function \cite{36} of
PSSTS. For this purpose, we start from the anti-normal ordering form of $\rho$
in Eq. (\ref{f7}). Recalling the integration formula converting an operator
$\hat{O}$ into its anti-normal ordering form\cite{37} with anti-normal
ordering $\vdots$ $\vdots$, i.e.,
\begin{equation}
\hat{O}=\int\frac{\mathtt{d}^{2}\beta}{\pi}\vdots\left\langle -\beta
\right\vert \hat{O}\left\vert \beta\right\rangle \exp\left(  \left\vert
\beta\right\vert ^{2}+\beta^{\ast}a-\beta a^{\dag}+a^{\dag}a\right)  \vdots,
\label{f18}%
\end{equation}
where $\left\vert \beta\right\rangle $ is a coherent state, one can obtain the
anti-normal ordering form of the squeezed thermal state $\rho_{s}$ by
substituting Eq.(\ref{f5}) into Eq. (\ref{f18}) and using the integration
formula$\ $(B7). Such as%

\begin{align}
\rho_{s}  &  =\frac{1}{\tau_{1}\tau_{2}}\int\frac{\mathtt{d}^{2}\beta}{\pi
}\vdots\exp\left\{  -A_{1}\left\vert \beta\right\vert ^{2}-\beta a^{\dag
}+\beta^{\ast}a+A_{2}\left(  \beta^{2}+\beta^{\ast2}\right)  +a^{\dag
}a\right\}  \vdots\nonumber\\
&  =\frac{1}{\sqrt{D}}\vdots\exp\left[  \left(  1-\frac{A_{1}}{E}\right)
a^{\dag}a+\frac{A_{2}}{E}\left(  a^{\dag2}+a^{2}\right)  \right]
\vdots\nonumber\\
&  =\frac{1}{\sqrt{D}}\vdots\exp\left[  \frac{2-\tau_{1}^{2}-\tau_{2}^{2}}%
{2D}a^{\dag}a+\frac{\tau_{1}^{2}-\tau_{2}^{2}}{4D}\left(  a^{\dag2}%
+a^{2}\right)  \right]  \vdots\label{f19}%
\end{align}
thus the anti-normal ordering form of $\rho$ in Eq. (\ref{f7}) is
\begin{equation}
\rho=\frac{C_{m}^{-1}}{\sqrt{D}}\vdots a^{m}\exp\left[  \frac{2-\tau_{1}%
^{2}-\tau_{2}^{2}}{2D}a^{\dag}a+\frac{\tau_{1}^{2}-\tau_{2}^{2}}{4D}\left(
a^{\dag2}+a^{2}\right)  \right]  a^{\dag m}\vdots, \label{f20}%
\end{equation}
which leads to the P-function $P\left(  \alpha\right)  $ of PSSTS,
\begin{equation}
P\left(  \alpha\right)  =C_{m}^{-1}\left\vert \alpha\right\vert ^{2m}%
P_{0}\left(  \alpha\right)  , \label{f21}%
\end{equation}
where $P_{0}\left(  \alpha\right)  $ is the P-function of STS,
\begin{equation}
P_{0}\left(  \alpha\right)  =\frac{1}{\sqrt{D}}\exp\left[  \frac{2-\tau
_{1}^{2}-\tau_{2}^{2}}{2D}\left\vert \alpha\right\vert ^{2}+\frac{\tau_{1}%
^{2}-\tau_{2}^{2}}{4D}\left(  \alpha^{\ast2}+\alpha^{2}\right)  \right]  .
\label{f22}%
\end{equation}
It is interesting in noticing that when $r=0$, Eq.(\ref{f22}) becomes
$P\left(  \alpha\right)  =\left\vert \alpha\right\vert ^{2m}e^{-\left\vert
\alpha\right\vert ^{2}/\bar{n}}/(\bar{n}C_{m}),$ which is just the P-function
of $m-$photon-subtraction thermal state which seems a new result. From
Eq.(\ref{f21}) one can see that the P-representation of density operator
$\rho$ can be expanded as
\begin{equation}
\rho=C_{m}^{-1}\int\frac{d^{2}z}{\pi}\left\vert z\right\vert ^{2m}P_{0}\left(
z\right)  \left\vert z\right\rangle \left\langle z\right\vert , \label{f26}%
\end{equation}
which is a non-Gaussian function due to the presence of $\left\vert
z\right\vert ^{2m}.$

\subsection{Q-function}

The Q-function is the absolute magnitude squared of the projection of a state
of the field onto a coherent state $\left\langle \alpha\right\vert $, defined
by%
\begin{equation}
Q\left(  \alpha,\alpha^{\ast}\right)  =\frac{1}{\pi}\left\langle
\alpha\right\vert \rho\left\vert \alpha\right\rangle . \label{f51}%
\end{equation}
Substituting Eq.(\ref{f26}) into (\ref{f51}), we can obtain%
\begin{equation}
Q\left(  \alpha,\alpha^{\ast}\right)  =R_{m}\left(  \alpha,\alpha^{\ast
}\right)  Q_{0}\left(  \alpha,\alpha^{\ast}\right)  , \label{f52}%
\end{equation}
where $Q_{0}\left(  \alpha,\alpha^{\ast}\right)  $ is the $Q$-function of
STS,
\begin{equation}
Q_{0}\left(  \alpha,\alpha^{\ast}\right)  =\frac{1}{\pi\allowbreak\tau_{1}%
\tau_{2}}\exp\left[  -\frac{\tau_{1}^{2}+\tau_{2}^{2}}{2\allowbreak\tau
_{1}^{2}\tau_{2}^{2}}\left\vert \alpha\right\vert ^{2}+\frac{\tau_{1}^{2}%
-\tau_{2}^{2}}{4\allowbreak\tau_{1}^{2}\tau_{2}^{2}}\left(  \alpha^{\ast
2}+\alpha^{2}\right)  \right]  , \label{f53}%
\end{equation}
which seems a new result not reported before, and $R_{m}\left(  \alpha
,\alpha^{\ast}\right)  $ is a factor generated from the photon-subtraction,
i.e.,%
\begin{equation}
R_{m}\left(  \alpha,\alpha^{\ast}\right)  =C_{m}^{-1}\sum_{l=0}^{m}%
\frac{\left(  m!\right)  ^{2}M^{m}\left(  2O\right)  ^{l}}{l!\left[  \left(
m-l\right)  !\right]  ^{2}}\left\vert H_{m-l}(-i\sqrt{M}(O\alpha^{\ast}%
+\alpha))\right\vert ^{2}, \label{f54}%
\end{equation}
where $M=[(2\bar{n}+1)\sinh2r]/[4\left(  \bar{n}^{2}+\left(  2\bar
{n}+1\right)  \allowbreak\cosh^{2}r\right)  ],$ and $O=2\bar{n}(\bar
{n}+1)/[\left(  2\bar{n}+1\right)  \sinh2r]$. Eq.(\ref{f52}) indicates that
the $Q$-function of PSSTS is also a non-Gaussian type due to the presence of
$R_{m}\left(  \alpha,\alpha^{\ast}\right)  $ and always positive since $O>0$.
In particular, when $m=0,R_{m}\left(  \alpha,\alpha^{\ast}\right)  =1,$ thus
$Q\left(  \alpha,\alpha^{\ast}\right)  =Q_{0}\left(  \alpha,\alpha^{\ast
}\right)  $, as expected.

\subsection{Wigner function}

Next, the P-function is applied to deduce the WF of PSSTS. The partial
negativity of WF is indeed a good indication of the highly nonclassical
character of the state. Therefore it is worth of obtaining the WF for any
states. The WF $W\left(  \alpha,\alpha^{\ast}\right)  $ associated with a
quantum state can be derived as follows\cite{28a}:
\begin{equation}
W\left(  \alpha,\alpha^{\ast}\right)  =\text{tr}[\rho\Delta\left(
\alpha,\alpha^{\ast}\right)  ],\text{ }\alpha=\left(  q+\mathtt{i}p\right)
/\sqrt{2}, \label{f23}%
\end{equation}
where $\Delta\left(  \alpha,\alpha^{\ast}\right)  $ is Wigner operator, whose
coherent state representation is
\begin{equation}
\Delta\left(  \alpha,\alpha^{\ast}\right)  =e^{2\left\vert \alpha\right\vert
^{2}}\int\frac{\mathtt{d}^{2}\beta}{\pi^{2}}\left\vert \beta\right\rangle
\left\langle -\beta\right\vert e^{2\left(  \alpha\beta^{\ast}-\alpha^{\ast
}\beta\right)  }, \label{f24}%
\end{equation}
where $\left\vert \beta\right\rangle =\exp(-\left\vert \beta\right\vert
^{2}/2+\beta a^{\dag})\left\vert 0\right\rangle $ is the coherent state. Using
the vacuum projector $\left\vert 0\right\rangle \left\langle 0\right\vert
=\colon e^{-a^{\dag}a}\colon$, and the IWOP technique \cite{28} one can put
Eq.(\ref{f24}) into its normal ordering form,
\begin{equation}
\Delta\left(  \alpha,\alpha^{\ast}\right)  =\frac{1}{\pi}\colon\exp\left[
-2\left(  a^{\dag}-\alpha^{\ast}\right)  \left(  a-\alpha\right)  \right]
\colon. \label{f25}%
\end{equation}
Thus substituting Eqs.(\ref{f22}), (\ref{f25}) and (\ref{f26}) into
Eq.(\ref{f23}), we can finally obtain the WF of PSSTS (see Appendix C),%

\begin{equation}
W\left(  \alpha,\alpha^{\ast}\right)  =F_{m}\left(  \alpha,\alpha^{\ast
}\right)  W_{0}\left(  \alpha,\alpha^{\ast}\right)  , \label{f27}%
\end{equation}
where $W_{0}\left(  \alpha,\alpha^{\ast}\right)  $ is the WF of STS,
\begin{equation}
W_{0}\left(  \alpha,\alpha^{\ast}\right)  =\frac{1}{\pi\allowbreak\left(
\allowbreak2\bar{n}+1\right)  \allowbreak}\exp\left[  -\frac{2\cosh2r}%
{2\bar{n}+1}\left\vert \alpha\right\vert ^{2}+\frac{\sinh2r}{\allowbreak
2\bar{n}+1}\left(  \alpha^{2}+\alpha^{\ast}{}^{2}\right)  \right]  ,
\label{f28}%
\end{equation}
and%
\begin{equation}
F_{m}\left(  \alpha,\alpha^{\ast}\right)  =\frac{\left(  m!\right)  ^{2}%
C_{m}^{-1}\sinh^{m}2r}{2^{2m}\left(  \allowbreak2\bar{n}+1\right)  ^{m}}%
\sum_{l=0}^{m}\frac{2^{2l}\left(  \bar{n}-\sinh^{2}r\right)  ^{l}}{l!\left[
\left(  m-l\right)  !\right]  ^{2}\sinh^{l}2r}\left\vert H_{m-l}\left(
\bar{\beta}\right)  \right\vert ^{2}, \label{f29}%
\end{equation}
where $\bar{\beta}=[2\alpha^{\ast}(\bar{n}-\sinh^{2}r)+\alpha\sinh
2r]/\{i[(2\bar{n}+1)\sinh2r]^{1/2}\}.$ Eq.(\ref{f27}) is the analytical
expression of WF for PSSTS, related to single-variable Hermite polynomials. It
is obvious that there does not exist negative region for WF in phase space
when $\bar{n}>\sinh^{2}r$ which is agreement with Eq.(28) in Ref.\cite{38}. In
particular, when $m=0,$ $F_{0}\left(  \alpha,\alpha^{\ast}\right)  =1,$
Eq.(\ref{f27}) becomes $W\left(  \alpha,\alpha^{\ast}\right)  =W_{0}\left(
\alpha,\alpha^{\ast}\right)  $; while for $r=0$, note $C_{m}=m!\bar{n}^{m}$,
$W_{0}\left(  \alpha,\alpha^{\ast}\right)  =e^{-2\left\vert \alpha\right\vert
^{2}/\allowbreak\left(  \allowbreak2\bar{n}+1\right)  }/\allowbreak\lbrack
\pi\left(  \allowbreak2\bar{n}+1\right)  ]$ (Eq.(30) in Ref.\cite{31}) and
$F_{m}\left(  \alpha,\alpha^{\ast}\right)  =\frac{1}{\left(  2\bar
{n}+1\right)  ^{m}}L_{m}\left(  -\frac{4\bar{n}}{2\bar{n}+1}\left\vert
\alpha\right\vert ^{2}\right)  $, Eq.(\ref{f27}) reduces to%
\begin{equation}
W\left(  \alpha,\alpha^{\ast}\right)  =\frac{1}{\pi\allowbreak\allowbreak
\left(  2\bar{n}+1\right)  ^{m+1}}e^{-\frac{2\left\vert \alpha\right\vert
^{2}}{2\bar{n}+1}}L_{m}\left(  -\frac{4\bar{n}\left\vert \alpha\right\vert
^{2}}{2\bar{n}+1}\right)  , \label{f30}%
\end{equation}
which corresponds to the WF of $m$-photon subtracted thermal state \cite{31},
and can be checked directly from Eq.(C6). In addition, for $m=1,$
(single-photon-subtracted squeezed thermal state (SPSSTS)), $C_{1}=B$
(\ref{f10}), the special WF of SPSSTS is%
\begin{equation}
W_{1}\left(  \alpha,\alpha^{\ast}\right)  =F_{1}\left(  \alpha,\alpha^{\ast
}\right)  W_{0}\left(  \alpha,\alpha^{\ast}\right)  , \label{f32}%
\end{equation}
where
\begin{align}
F_{1}\left(  \alpha,\alpha^{\ast}\right)   &  =\frac{1}{\left(  \allowbreak
2\bar{n}+1\right)  B}\left(  \left(  2\bar{n}+1\right)  \left\vert \bar
{\alpha}\right\vert ^{2}+\bar{n}-\sinh^{2}r\right) \nonumber\\
&  =\frac{\left\vert 2\alpha^{\ast}\left(  \bar{n}-\sinh^{2}r\right)
+\alpha\sinh2r\right\vert ^{2}}{\left(  \allowbreak2\bar{n}+1\right)  ^{2}%
B}+\frac{\bar{n}-\sinh^{2}r}{\left(  \allowbreak2\bar{n}+1\right)  B}.
\label{f31}%
\end{align}

Noting $B>0$, thus from Eq.(\ref{f32}) one can see that when the factor
$F_{1}\left(  \alpha,\alpha^{\ast}\right)  <0,$ the WF of SPSSTS has its
negative distribution in phase space. This indicates that the WF of SPSSTS
always has the negative values under the condition: $\bar{n}<\sinh^{2}r$\ at
the phase space center $\alpha=0,$ which is similar to the case of
single-photon-subtracted squeezed vacuum \cite{2d,2f}.

\begin{figure}[ptb]
\label{Fig3}
\centering\includegraphics[width=14cm]{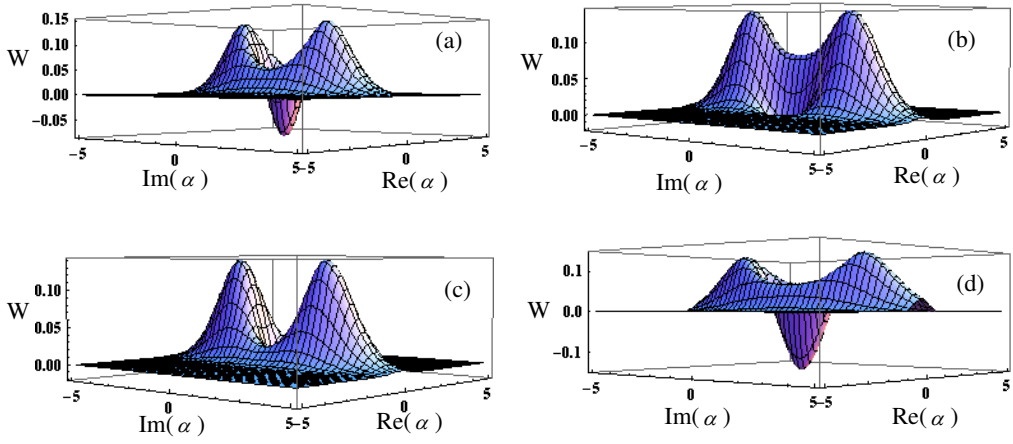}\caption{{\protect\small (Color
online) Wigner function distributions }${\protect\small W}\left(
\alpha,\alpha^{\ast}\right)  ${\protect\small of PASSTS} {\protect\small for
different (}$\bar{n},r${\protect\small ) and }$m$ {\protect\small values (a)
}$\bar{n}=0.1,r=0.5,m=1;${\protect\small (b) }$\bar{n}=0.1,r=0.5,m=2;$%
{\protect\small (c) }$\bar{n}=0.2,r=0.5,m=1;${\protect\small (d) }$\bar
{n}=0.1,r=0.8,m=1.$}%
\end{figure}

Using Eq.(\ref{f27}), the WFs of PSSTS are depicted in Fig.3 for several
different values of $\bar{n},r$ and $m$ in phase space. It is easy to see that
the the WF is non-Gaussian in phase space. As an evidence of nonclassicality
of the state, squeezing in one of the quadratures is clear in the plots (see
Figs.3(a) and 3(d)). In addition, we can clearly see that there is some
negative region of WF, which is another evidence of nonclassicality of the
state, and that the negative region of WF gradually disappears as the $\bar
{n}$ (or the temperature) increases for given $r$ and $m$ (see Fig.3(a) and
(c)). Furthermore, for a larger squeezing, the WF shows a smaller minimum
negative value at the center of phase space (Figs.3(a) and 3(d)). For
two-photon subtracted case, the WF presents two positive peaks and two
negative peaks, different from the case of single-photon subtracted case.

\section{Decoherence of PSSTS in thermal environment}

When the $m$-PSSTS evolves in the thermal channel, the evolution of the
density matrix can be described by master equation \cite{r22}%
\begin{equation}
\frac{d\rho}{dt}=\kappa\left(  \mathfrak{N}+1\right)  \left(  2a\rho
a^{\dagger}-a^{\dagger}a\rho-\rho a^{\dagger}a\right)  +\kappa\mathfrak{N}%
\left(  2a^{\dagger}\rho a-aa^{\dagger}\rho-\rho aa^{\dagger}\right)  ,
\label{f33}%
\end{equation}
where $\kappa$ represents the dissipative coefficient and $\mathfrak{N}$
denotes the average thermal photon number of the environment. When
$\mathfrak{N}=0,$ Eq.(\ref{f33}) reduces to the master equation describing the
photon-loss channel. The evolution of the WF is governed by the following
integration equation \cite{oc},
\begin{equation}
W\left(  \zeta,\zeta^{\ast},t\right)  =\frac{2}{\left(  2\mathfrak{N}%
+1\right)  \mathcal{T}}\int\frac{d^{2}\alpha}{\pi}W\left(  \alpha,\alpha
^{\ast},0\right)  e^{-2\frac{\allowbreak\left\vert \zeta-\alpha e^{-\kappa
t}\right\vert ^{2}}{\left(  2\mathfrak{N}+1\right)  \mathcal{T}}}, \label{f34}%
\end{equation}
where $W\left(  \alpha,\alpha^{\ast},0\right)  $ is the WF of the initial
state, and $\mathcal{T}=1-e^{-2\kappa t}$. Eq.(\ref{f34}) is just the
evolution formula of WF in thermal channel. Thus the WF at evolving time may
be obtained by performing the integration with an initial value.

Substituting Eqs.(\ref{f27})-(\ref{f29}) into (\ref{f34}), and using Eq.(B7)
we finally obtain the evolution of WF for PSSTS in thermal environment (see
Appendix D, in a similar way to deriving Eq.\textbf{(\ref{f8})}),
\begin{equation}
W\left(  \zeta,\zeta^{\ast},t\right)  =F_{m}\left(  \zeta,\zeta^{\ast
},t\right)  W_{0}\left(  \zeta,\zeta^{\ast},t\right)  , \label{f35}%
\end{equation}
where $W_{0}\left(  \zeta,\zeta^{\ast},t\right)  $ is the WF of squeezed
thermal state in thermal channel,
\begin{align}
W_{0}\left(  \zeta,\zeta^{\ast},t\right)   &  =\frac{1/\allowbreak\left(
\allowbreak2\bar{n}+1\right)  }{\pi\left(  2\mathfrak{N}+1\right)
\mathcal{T}\sqrt{G}}\exp\left[  -\Delta_{2}\left\vert \zeta\right\vert
^{2}+\frac{\allowbreak g_{2}g_{3}^{2}}{4G}\left(  \zeta^{2}+\zeta^{\ast
2}\right)  \right]  ,\label{f36}\\
F_{m}\left(  \zeta,\zeta^{\ast},t\right)   &  =C_{m}^{-1}\sum_{l=0}^{m}%
\frac{\left(  m!\right)  ^{2}\chi^{l}\Delta_{1}^{m-l}}{l!\left[  \left(
m-l\right)  !\right]  ^{2}}\left\vert H_{m-l}\left[  \omega/(2i\sqrt
{\Delta_{1}})\right]  \right\vert ^{2}, \label{f37}%
\end{align}
and%
\begin{equation}
g_{0}=\frac{\cosh2r}{2\bar{n}+1},\text{ }g_{1}=\frac{\bar{n}-\sinh^{2}r}%
{2\bar{n}+1},\text{ }g_{2}=\frac{\sinh2r}{2\bar{n}+1},\text{ }g_{3}%
=\frac{2e^{-\kappa t}}{\left(  2\mathfrak{N}+1\right)  \mathcal{T}},
\label{f38}%
\end{equation}
as well as%
\begin{align}
G  &  =\left(  g_{0}+g_{3}e^{-\kappa t}/2\right)  ^{2}-g_{2}^{2},\nonumber\\
\omega &  =\frac{2e^{-\kappa t}}{2\mathfrak{N}\mathcal{T}+1}\left(  \chi
\zeta+2\Delta_{1}\zeta\allowbreak^{\ast}\right)  ,\nonumber\\
\Delta_{1}  &  =\frac{g_{2}}{4G}\left(  1+g_{3}e^{-\kappa t}/2\right)
^{2},\label{f39}\\
\Delta_{2}  &  =\frac{2\allowbreak}{\left(  2\mathfrak{N}+1\right)
\mathcal{T}}-\allowbreak\frac{g_{3}^{2}}{2G}\left(  g_{0}+g_{3}e^{-\kappa
t}/2\right)  ,\nonumber\\
\chi &  =\frac{1+g_{3}e^{-\kappa t}/2}{2G}\left[  g_{0}\allowbreak+g_{1}%
g_{3}e^{-\kappa t}-1/\left(  2\bar{n}+1\right)  ^{2}\allowbreak\right]
.\nonumber
\end{align}
It is noted that, at the initial time ($t=0$), $\left(  2\mathfrak{N}%
+1\right)  T\sqrt{G}\rightarrow1,\Delta_{2}\rightarrow2g_{0},\frac{\allowbreak
g_{2}g_{3}^{2}}{4G}\rightarrow\allowbreak g_{2},\Delta_{1}\rightarrow\frac
{1}{4}g_{2},\chi\rightarrow\allowbreak g_{1},$ $\omega\rightarrow2g_{1}%
\zeta+g_{2}\zeta\allowbreak^{\ast},$ Eqs.(\ref{f36}) and (\ref{f37}) just
reduce to Eqs.(\ref{f28}) and (\ref{f29}), respectively, i.e., the WF of
PSSTS. In addition, for the case of $m=1$, corresponding to the case of
SPSSTS, Eq.(\ref{f37}) just becomes%
\begin{equation}
F_{1}\left(  \zeta,\zeta^{\ast},t\right)  =C_{1}^{-1}\left(  \left\vert
\omega\right\vert ^{2}+\chi\right)  , \label{f41}%
\end{equation}
from which one can see that when the factor $F_{1}\left(  \zeta,\zeta^{\ast
},t\right)  <0,$ the WF of SPSSTS in thermal channel has its negative
distribution in phase space. At the phase space center $\zeta=0,$ the WF of
SPSSTS always has the negative values when $\chi<0$, leading to the following
condition:
\begin{equation}
\kappa t<\kappa t_{c}=\frac{1}{2}\ln\left[  1-\frac{2\bar{n}+1}{2\mathfrak{N}%
+1}\frac{\bar{n}-\sinh^{2}r}{\bar{n}\cosh2r+\sinh^{2}r}\right]  , \label{f42}%
\end{equation}
which\ implies that the threshold value $\kappa t_{c}$ is dependent not only
on the average number $\mathfrak{N}$ of environment, but also on the average
number $\bar{n}$ of thermal state and the squeezing parameter $r$ (a result
different from other discussions about the threshold value $\kappa t_{c}$ in
thermal channel \cite{2d,2f}). The WF of SPSSTS is always positive in the
whole phase space when $\kappa t\ $exceeds the threshold value $\kappa t_{c}$.
Actually, Eq.(\ref{f42}) is also true for the case with any number ($m$)
photon-subtraction (see Eq.(\ref{f37})). From Eq.(\ref{f42}) one can clarify
how the thermal noise $\left(  \bar{n},\mathfrak{N}\right)  $ shortens the
threshold value of the decay time.

Using Eq. (\ref{f35}) we present the time-evolution of WF at different times
scales in Fig.4. From Fig.4, one can see clearly that the partial negative
region of WF gradually diminishes. At long times $\kappa t\rightarrow\infty$,
one has $\omega\rightarrow0,\chi\rightarrow\bar{n}\cosh2r+\sinh^{2}%
r,\Delta_{1}\rightarrow\frac{1}{4}\left(  2\bar{n}+1\right)  \allowbreak
\sinh2r$ and $H_{m}(0)=\left(  -1\right)  ^{k}m!\delta_{m,2k}/k!.$ Thus
\begin{equation}
W\left(  \zeta,\zeta^{\ast},\infty\right)  =\frac{1}{\pi\left(  2\mathfrak{N}%
+1\right)  }e^{-\frac{2\allowbreak\left\vert \zeta\right\vert ^{2}%
}{2\mathfrak{N}+1}}, \label{f40}%
\end{equation}
which is independent of photon-subtraction number $m$ and corresponds to
thermal states with mean thermal photon number $\mathfrak{N}$. This implies
that the system reduces to thermal state after a long time interaction with
the environment. Eq.(\ref{f40}) denotes a Gaussian distribution. Thus the
thermal noise causes the absence of the partial negative of the WF if the
decay time $\kappa t$ exceeds a threshold value. In addition, from Fig.4, it
is found that the SPSSTS is similar to a Schrodinger cat state.

\begin{figure}[ptb]
\label{Fig4}
\centering\includegraphics[width=14cm]{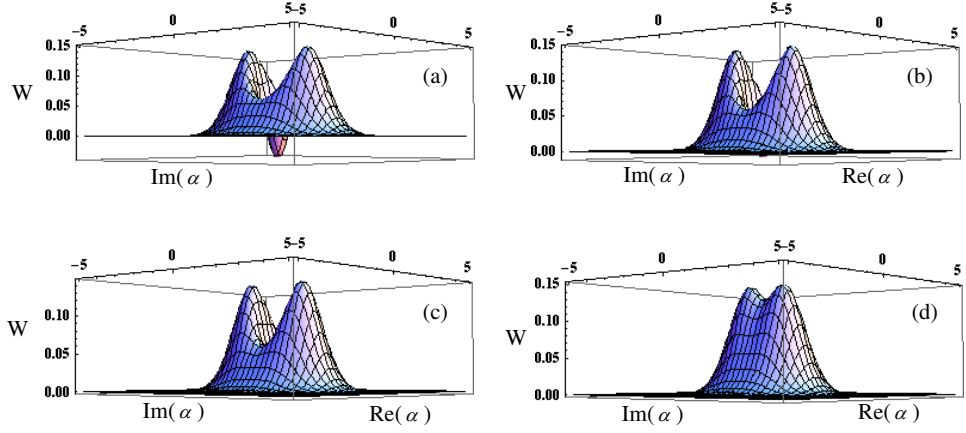}\caption{{\protect\small (Color
online) Wigner function }${\protect\small W}\left(  \alpha,\alpha^{\ast
},t\right)  ${\protect\small of SPSSTS} {\protect\small for }$r=0.3,\bar
{n}=0.05$ {\protect\small and different }$r${\protect\small and }$\kappa
t${\protect\small : (a) }$\kappa t=0.05;${\protect\small (b) }$\kappa
t=0.1;${\protect\small (c) }$\kappa t=0.2;${\protect\small (d) }$\kappa
t=0.5.$}%
\end{figure}

\section{Fidelity as a non-Gaussianity measure for PSSTS}

Recently, some quantitative measures to assess non-Gaussianity are proposed
\cite{39,40}. A non-Gaussianity measure may serve as a guideline to quantify
the non-Gaussian states. Therefore, it is of interest to evaluate the degree
of the resulting non-Gaussianity and assess this operation as a resource to
obtain non-Gaussian states starting from Gaussian ones. Here, we examine the
fidelity between the PSSTS $\rho_{s}$ and the STS $\rho$. Since the STS can be
considered as a generalized Gaussian state, the fidelity may be seen as a
non-Gaussianity measure able to quantify the non-Gaussian character of a
quantum state. In order to quantify the non-Gaussian character of the PSSTS,
we introduce the fidelity by defining%
\begin{equation}
\mathfrak{F}=\mathtt{tr}\left(  \rho_{s}\rho\right)  /\mathtt{tr}\left(
\rho_{s}^{2}\right)  , \label{f43}%
\end{equation}
where\ $\rho_{s}$ and $\rho$ are the squeezed thermal state (a generalized
Gaussian state) and the PSSTS, respectively. Obviously, when
photon-subtraction number $m=0,$ leading to $\rho=\rho_{s}$, then
$\mathfrak{F}=1$ which means that $\rho$ is a Gaussian state described by
$\rho_{s}$.

Using Eqs.(\ref{f1}) and (\ref{f2}), one has
\begin{equation}
\mathtt{tr}\left(  \rho_{s}^{2}\right)  =\mathtt{tr}\left(  \rho_{c}%
^{2}\right)  =\frac{1}{2\bar{n}+1}. \label{f44}%
\end{equation}
On the other hand, the fidelity ($\mathtt{tr}\left(  \rho_{s}\rho\right)  $)
can then be calculated as the overlap between the two WFs:%
\begin{equation}
\mathtt{tr}\left(  \rho_{s}\rho\right)  =4\pi\int d^{2}\alpha W_{0}\left(
\alpha,\alpha^{\ast}\right)  W_{\rho}\left(  \alpha,\alpha^{\ast}\right)  ,
\label{f45}%
\end{equation}
where $W_{0}\left(  \alpha,\alpha^{\ast}\right)  $ is the WF of squeezed
thermal state $\rho_{s}$. Using Eq.(\ref{f27}) we may express Eq.(\ref{f45})
as%
\begin{equation}
\mathtt{tr}\left(  \rho_{s}\rho\right)  =4\pi\int F_{m}\left(  \alpha
,\alpha^{\ast}\right)  W_{0}^{2}\left(  \alpha,\alpha^{\ast}\right)
d^{2}\alpha. \label{f46}%
\end{equation}
Then employing Eqs.(\ref{f28}) and (C6), similarly to Eq.(\ref{f8}), Eq.
(\ref{f46}) may rewritten as (see Appendix E)%
\begin{equation}
\mathtt{tr}\left(  \rho_{s}\rho\right)  =\frac{m!B_{2}^{m/2}}{\allowbreak
\allowbreak\left(  2\bar{n}+1\right)  C_{m}}P_{m}\left(  B_{1}/\sqrt{B_{2}%
}\right)  , \label{f47}%
\end{equation}
where $P_{m}\left(  x\right)  $ is the Legendre polynomial with
\begin{equation}
B_{1}=\frac{\allowbreak\bar{n}\left(  \bar{n}+1\right)  }{2\bar{n}+1}%
\cosh2r,\text{ }B_{2}=\frac{\bar{n}^{2}\left(  \bar{n}+1\right)  ^{2}}{\left(
2\bar{n}+1\right)  ^{2}}-\sinh^{2}r\cosh^{2}r. \label{f48}%
\end{equation}
Thus the fidelity (\ref{f43}) for the PSSTS is given by%
\begin{equation}
\mathfrak{F}=\frac{m!}{C_{m}}B_{2}^{m/2}P_{m}\left(  B_{1}/\sqrt{B_{2}%
}\right)  =\left(  \frac{B_{2}}{D}\right)  ^{m/2}\frac{P_{m}\left(
B_{1}/\sqrt{B_{2}}\right)  }{P_{m}\left(  B/\sqrt{D}\right)  }, \label{f49}%
\end{equation}
which is an analytical expression for the fidelity between PSSTS and SSTS. We
see that when $m=0$ (the case of no photon-subtraction), $\mathfrak{F}=1$;
while for $m=1$ (the case of SPSSTS), Eq.(\ref{f49}) reduces to%
\begin{equation}
\mathfrak{F=}\frac{\allowbreak\bar{n}\left(  \bar{n}+1\right)  \cosh
2r}{\left(  2\bar{n}+1\right)  \left(  \sinh^{2}r+\bar{n}\cosh2r\right)  }.
\label{f50}%
\end{equation}

In Fig.5, we plot the fidelity between PSSTS and STS as the function of
squeezing parameter $r$\ for different photon-subtraction number $m.$ From
Fig. 5 one can see that the fidelity decreases monotonously with the increment
of both photon-subtraction number $m$ and the squeezing parameter $r$, as expected.

\begin{figure}[ptb]
\label{Fig5}
\centering\includegraphics[width=10cm]{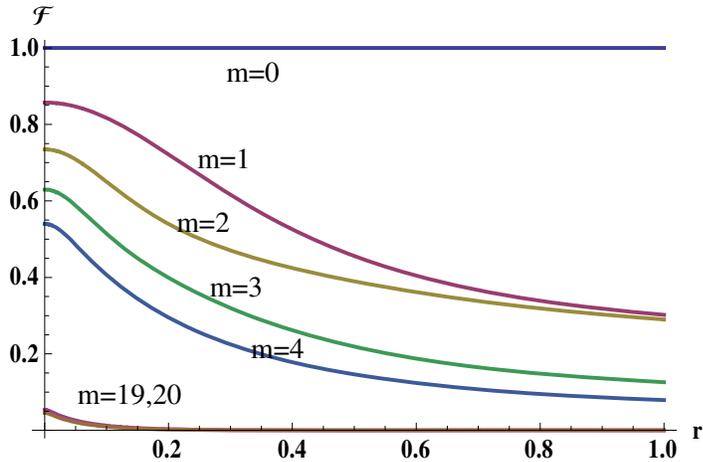}\caption{{\protect\small (Color
online) The fidelity between PSSTS and squeezed thermal state as the function
of squeezing parameter }$r${\protect\small \ for different photon-subtraction
number }$m=0,1,2,3,4,19,20.(\bar{n}=0.2)$. {\protect\small The cases of
}${\protect\small m=}${\protect\small 19 and 20 are not identical, but they
are almost overlap each other, which can not be seen clearly from figure due
to the use of thick style for line.}}%
\end{figure}

\section{Conclusions and Remarks}

In summary, we investigate the nonclassicality photon-subtracted squeezed
thermal state (PSSTS) and its decoherence in thermal channel with average
thermal photon number $\mathfrak{N}$\ and dissipative coefficient $\kappa$.
For arbitrary number PSSTS, we have, for the first time, obtained an
analytical express for the normalization factor, which turns out to be a
Legendre polynomial of squeezing parameter $r$\ and average photon number
$\bar{n}$ of thermal state, a remarkable result. Based on Legendre
polynomials' behavior the nonclassical properties of the field, such as
Mandel's $Q$-parameter and photon number distribution, are also derived
analytically, Furthermore, the nonclassicality of PSSTS is discussed in terms
of the negativity of WF after deriving the explicit expression of WF, which
implies the highly nonclassical properties of quantum states. It is shown that
the WF of single PSSTS always has negative values if $\bar{n}<\sinh^{2}r$\ at
the phase space center. Then the decoherence of PSSTS in thermal channel is
also demonstrated according to the compact expression for the WF. It is found
that the threshold value of the decay time corresponding to the transition of
the WF from partial negative to completely positive definite is obtained at
the center of the phase space, which is dependent not only on the average
number $\mathfrak{N}$ of environment, but also on the average number $\bar{n}$
of thermal state and the squeezing parameter $r$. We show that the WF for
single PSSTS has always negative value if the decay time $\kappa t<\frac{1}%
{2}\ln\{1-(2\bar{n}+1)(\bar{n}-\sinh^{2}r)/[(2\mathfrak{N}+1)(\bar{n}%
\cosh2r+\sinh^{2}r)]\}$. A non-Gaussianity measure may serve as a guideline to
quantify them for the class of non-Gaussian states, where the fidelity
decreases monotonously with the increment of both photon-subtraction number
$m$ and the squeezing parameter $r$.

In addition, Mandel's $Q$ parameter does not always indicate a negative value
for non-classical state. In fact, for the photon-subtracted squeezed states by
even number, this parameter is positive. Thus the negativity of Q parameter is
a sufficient condition to distinguish non-classical state from classical one.
While for photon subtracted squeezed states by odd number, the negativity of
single photon subtracted case is noticeable. To compare further
non-classicality of quantum states for different number subtracted case, the
measures based on the volume of the negative part of the Wigner
function\cite{40a}, on the nonclassical depth \cite{40b} and on the
entanglement potential \cite{40c} may be other alternative methods.
Non-classical state introduced in this work will maybe used in combination
with other non-classical states such as entangled states.

On the other hand, we should mention that for a photon-subtracted squeezed
state generated with some realistic probability, its non-classicality, in
particular, its non-Gaussianity would not be always superior to the input
Gaussian state. For example, photon-subtracted two-mode squeezed vacuum state
has more entanglement than initial two-mode squeezed state in not so strong
squeezing parameter; while for strong squeezing region, its superiority
disappears \cite{Ki}. Entanglement evaluation investigation for
photon-subtracted two-mode squeezed thermal state is a future problem.

\textbf{Acknowledgments }Work supported by the the National Natural Science
Foundation of China under Grant Nos.10775097, and a grant from the Key
Programs Foundation of Ministry of Education of China (No. 210115), and the
Research Foundation of the Education Department of Jiangxi Province of China
(No. GJJ10097).

\textbf{APPENDIX A: Derivation of Eq.(\ref{f5})}

Using the operator identity (\ref{f3}) and noticing the single-mode squeezing
operator yields the transformations,
\begin{equation}
S\left(  r\right)  QS^{\dag}\left(  r\right)  =e^{-r}Q,\text{ \ }S\left(
r\right)  PS^{\dag}\left(  r\right)  =e^{r}P, \tag{A1}%
\end{equation}
one has
\begin{equation}
\rho_{s}=(1-e^{\sigma})S(r)e^{\sigma a^{\dagger}a}S^{\dagger}(r)=\frac
{2(1-e^{\sigma})}{e^{\sigma}+1}%
%TCIMACRO{\QATOP{:}{:}}%
%BeginExpansion
\genfrac{}{}{0pt}{}{:}{:}%
%EndExpansion
\exp\left\{  \frac{e^{\sigma}-1}{e^{\sigma}+1}\left(  e^{-2r}Q^{2}+e^{2r}%
P^{2}\right)  \right\}
%TCIMACRO{\QATOP{\colon}{\colon}}%
%BeginExpansion
\genfrac{}{}{0pt}{}{\colon}{\colon}%
%EndExpansion
, \tag{A2}%
\end{equation}
which is still in Weyl ordering, in deriving (A2) we have used the Weyl
ordering invariance under similarity transformations (\ref{f4}). According to
the definition of Weyl correspondence rule \cite{41}, i.e., the classical Weyl
function $f\left(  q,p\right)  $ of operator $\rho_{s}$ can be given by
replacing the $Q$ and by $q$ and $p$ in its Weyl ordered form, respectively,
\begin{equation}
f\left(  q,p\right)  =\frac{2(1-e^{\sigma})}{e^{\sigma}+1}\exp\left\{
\frac{e^{\sigma}-1}{e^{\sigma}+1}\left(  e^{-2r}q^{2}+e^{2r}p^{2}\right)
\right\}  , \tag{A3}%
\end{equation}
then using the relation between $\rho_{s}$ and Wigner operator $\Delta\left(
q,p\right)  ,$ i.e., operator $\rho_{s}$ can be expanded in terms of
$\Delta\left(  q,p\right)  $,
\begin{equation}
\rho_{s}=\int_{-\infty}^{\infty}dqdpf\left(  q,p\right)  \Delta\left(
q,p\right)  , \tag{A4}%
\end{equation}
where the normal ordering form of $\Delta\left(  q,p\right)  $ is given by%
\begin{equation}
\Delta\left(  q,p\right)  =\frac{1}{\pi}\colon\exp\left[  -\left(  q-Q\right)
^{2}-\left(  p-P\right)  ^{2}\right]  \colon\tag{A5}%
\end{equation}
one can see that
\begin{align}
\rho_{s}  &  =\frac{2(1-e^{\sigma})}{\pi\left(  e^{\sigma}+1\right)  }%
\int_{-\infty}^{\infty}dqdp\exp\left\{  \frac{e^{\sigma}-1}{e^{\sigma}%
+1}\left(  e^{-2r}q^{2}+e^{2r}p^{2}\right)  \right\} \nonumber\\
&  \times\colon\exp\left[  -\left(  q-Q\right)  ^{2}-\left(  p-P\right)
^{2}\right]  \colon\nonumber\\
&  =\text{Eq.(\ref{f5}).} \tag{A6}%
\end{align}
thus we complete the proof of Eq.(\ref{f5}).

\textbf{APPENDIX B: Deduction of Eq.(\ref{f8})}

Using the completeness relation and $\rho_{s}^{\prime}s$ normal ordering form
in (\ref{f5}), as well as the overlap of coherent state,
\begin{equation}
\left\langle \beta\right\vert \left.  \alpha\right\rangle =\exp\left[
-\frac{1}{2}\left\vert \alpha\right\vert ^{2}-\frac{1}{2}\left\vert
\beta\right\vert ^{2}+\beta^{\ast}\alpha\right]  , \tag{B1}%
\end{equation}
we have%
\begin{align}
C_{m}  &  =\frac{1}{\tau_{1}\tau_{2}}\mathtt{Tr}\left\{  a^{m}\int\frac
{d^{2}\alpha d^{2}\beta}{\pi^{2}}\left\vert \alpha\right\rangle \left\langle
\alpha\right\vert \colon\exp\left[  -\frac{Q^{2}}{2\tau_{1}^{2}}-\frac{P^{2}%
}{2\tau_{2}^{2}}\right]  \colon\left\vert \beta\right\rangle \left\langle
\beta\right\vert a^{\dag m}\right\} \nonumber\\
&  =\frac{1}{\tau_{1}\tau_{2}}\int\frac{d^{2}\alpha d^{2}\beta}{\pi^{2}}%
\alpha^{m}\beta^{\ast m}\exp\left[  -\left\vert \alpha\right\vert
^{2}-\left\vert \beta\right\vert ^{2}+\beta^{\ast}\alpha+A_{1}\beta
\alpha^{\ast}+A_{2}\left(  \beta^{2}+\alpha^{\ast2}\right)  \right]
\nonumber\\
&  =\frac{1}{\tau_{1}\tau_{2}}\frac{\partial^{2m}}{\partial k^{m}\partial
s^{m}}\int\frac{d^{2}\alpha d^{2}\beta}{\pi^{2}}\exp\left[  -\left\vert
\alpha\right\vert ^{2}+\left(  \beta^{\ast}+k\right)  \alpha+A_{1}\beta
\alpha^{\ast}+A_{2}\alpha^{\ast2}\right] \nonumber\\
&  \times\left.  \exp\left[  -\left\vert \beta\right\vert ^{2}+\beta^{\ast
}s+A_{2}\beta^{2}\right]  \right\vert _{s=k=0}\nonumber\\
&  =\frac{1}{\tau_{1}\tau_{2}}\frac{\partial^{2m}}{\partial k^{m}\partial
s^{m}}e^{A_{2}k^{2}}\int\frac{d^{2}\beta}{\pi}\exp\left[  -\left(
1-A_{1}\right)  \left\vert \beta\right\vert ^{2}+kA_{1}\beta+\left(
s+2A_{2}k\right)  \beta^{\ast}+A_{2}\left(  \beta^{2}+\beta^{\ast2}\right)
\right]  _{s=k=0}\nonumber\\
&  =\frac{1}{\tau_{1}\tau_{2}\sqrt{A_{3}}}\frac{\partial^{2m}}{\partial
k^{m}\partial s^{m}}\exp\left[  \left(  k^{2}+s^{2}\right)  A+Bks\right]
_{s=k=0}, \tag{B2}%
\end{align}
where
\begin{align}
A_{1}  &  =1-\frac{1}{2\tau_{2}^{2}}-\frac{1}{2\tau_{1}^{2}},A_{2}=\frac
{1}{4\tau_{2}^{2}}-\frac{1}{4\tau_{1}^{2}}>0,\tag{B3}\\
A_{3}  &  =\left(  1-A_{1}\right)  ^{2}-4A_{2}^{2}=\frac{1}{\tau_{1}^{2}%
\tau_{2}^{2}},\tag{B4}\\
A  &  =\frac{A_{2}}{A_{3}}=\allowbreak\frac{1}{4}\left(  \tau_{1}^{2}-\tau
_{2}^{2}\right)  =\allowbreak\frac{2\bar{n}+1}{4}\sinh2r>0,\tag{B5}\\
B  &  =\frac{A_{1}-A_{1}^{2}+4A_{2}^{2}}{A_{3}}=\allowbreak\frac{1}{2}\left(
\tau_{1}^{2}+\tau_{2}^{2}\right)  -1\nonumber\\
&  =\frac{1}{2}\left[  \left(  2\bar{n}+1\right)  \cosh2r-1\right]  >0,
\tag{B6}%
\end{align}
and using the integration formula \cite{PP}
\begin{equation}
\int\frac{d^{2}z}{\pi}\exp\left\{  \zeta\left\vert z\right\vert ^{2}+\xi
z+\eta z^{\ast}+fz^{2}+gz^{\ast2}\right\}  =\frac{1}{\sqrt{\zeta^{2}-4fg}}%
\exp\left\{  \frac{-\zeta\xi\eta+\xi^{2}g+\eta^{2}f}{\zeta^{2}-4fg}\right\}  ,
\tag{B7}%
\end{equation}
whose convergent condition is Re$\left(  \zeta\pm f\pm g\right)  <0$
and$\ \mathtt{Re}\left(  \frac{\zeta^{2}-4fg}{\zeta\pm f\pm g}\right)  <0$ and
noting that%
\begin{align}
&  \frac{\partial^{2m}}{\partial t^{m}\partial\tau^{m}}\left.  \exp\left(
-t^{2}-\tau^{2}+2x\tau t\right)  \right\vert _{t,\tau=0}\nonumber\\
&  =\sum_{n,l,k=0}^{\infty}\frac{\left(  -\right)  ^{n+l}}{n!l!k!}\left(
2x\right)  ^{k}\left.  \frac{\partial^{2m}}{\partial t^{m}\partial\tau^{m}%
}\tau^{2n+k}t^{2l+k}\right\vert _{t,\tau=0}\nonumber\\
&  =2^{m}m!\sum_{n=0}^{\left[  m/2\right]  }\frac{m!}{2^{2n}\left(  n!\right)
^{2}\left(  m-2n\right)  !}x^{m-2n}, \tag{B8}%
\end{align}
one rewritten Eq. (B2) as%
\begin{align}
C_{m}  &  =\left(  -A\right)  ^{m}\frac{\partial^{2m}}{\partial k^{m}\partial
s^{m}}\exp\left[  -k^{2}-s^{2}-\frac{B}{A}ks\right]  _{s=k=0}\nonumber\\
&  =\left(  -A\right)  ^{m}2^{m}m!\sum_{n=0}^{\left[  m/2\right]  }%
\frac{m!\left(  \frac{-B}{2A}\right)  ^{m-2n}}{2^{2n}\left(  n!\right)
^{2}\left(  m-2n\right)  !}. \tag{B9}%
\end{align}
Recalling the newly found expression of Legendre polynomial (its equivalence
to the well-known Legendre polynomial's ($P_{m}\left(  x\right)  $) expression
is \cite{r33}
\begin{equation}
x^{m}\sum_{{l}=0}^{\left[  m/2\right]  }\frac{m!}{2^{2{l}}\left(  {l}!\right)
^{2}\left(  m-2{l}\right)  !}\left(  1-\frac{1}{x^{2}}\right)  ^{{l}}%
=P_{m}\left(  x\right)  , \tag{B10}%
\end{equation}
we derive the compact form for $C_{m}$,%

\begin{align}
C_{m}  &  =m!B^{m}\sum_{n=0}^{\left[  m/2\right]  }\frac{m!}{2^{2n}\left(
n!\right)  ^{2}\left(  m-2n\right)  !}\left(  \frac{4A^{2}}{B^{2}}\right)
^{n}\nonumber\\
&  =m!D^{m/2}P_{m}\left(  B/\sqrt{D}\right)  , \tag{B11}%
\end{align}
where
\begin{equation}
D=\allowbreak B^{2}-4A^{2}=\bar{n}^{2}-\left(  2\bar{n}+1\right)  \sinh^{2}r.
\tag{B12}%
\end{equation}
Eq.(B11) indicates that the normalization factor $C_{m}$ is just related to
Legendre polynomial.

Combining Eqs.(B8) and (B10), On the other hand, one can derive a new formula
for Legendre polynomial, i.e.,
\begin{equation}
\frac{\partial^{2m}}{\partial t^{m}\partial\tau^{m}}\left.  \exp\left(
-t^{2}-\tau^{2}+\frac{2x\tau t}{\sqrt{x^{2}-1}}\right)  \right\vert
_{t,\tau=0}=\frac{2^{m}m!}{\left(  x^{2}-1\right)  ^{m/2}}P_{m}\left(
x\right)  . \tag{B13}%
\end{equation}

\textbf{APPENDIX\ C:} \textbf{Derivation of WF (\ref{f8}) for PSSTS}

According to Eqs.(\ref{f23}), (\ref{f25}) and (\ref{f26}), we have%
\begin{align}
W\left(  \alpha,\alpha^{\ast}\right)   &  =C_{m}^{-1}\text{tr}[\int\frac
{d^{2}z}{\pi}\left\vert z\right\vert ^{2m}P_{0}\left(  z\right)  \left\vert
z\right\rangle \left\langle z\right\vert \Delta\left(  \alpha,\alpha^{\ast
}\right)  ]\nonumber\\
&  =\frac{C_{m}^{-1}}{\pi}\int\frac{d^{2}z}{\pi}\left\vert z\right\vert
^{2m}P_{0}\left(  z\right)  \exp\left[  -2\left(  z^{\ast}-\alpha^{\ast
}\right)  \left(  z-\alpha\right)  \right] \nonumber\\
&  =\frac{C_{m}^{-1}e^{-2\left\vert \alpha\right\vert ^{2}}}{\pi\sqrt{D}}%
\int\frac{d^{2}z}{\pi}\left\vert z\right\vert ^{2m}\exp\left[  -g\left\vert
z\right\vert ^{2}+2\alpha^{\ast}z+2\alpha z^{\ast}+\frac{\tau_{-}}{4D}\left(
z^{\ast2}+z^{2}\right)  \right] \nonumber\\
&  =\frac{C_{m}^{-1}e^{-2\left\vert \alpha\right\vert ^{2}}}{\pi\sqrt{D}}%
\frac{\partial^{2m}}{\partial k^{m}\partial t^{m}}\int\frac{d^{2}z}{\pi}%
\exp\left[  -g\left\vert z\right\vert ^{2}+\left(  2\alpha+k\right)  z^{\ast
}+\left(  2\alpha^{\ast}+t\right)  z+\frac{\tau_{-}}{4D}\left(  z^{\ast
2}+z^{2}\right)  \right]  _{k=t=0}, \tag{C1}%
\end{align}
where
\begin{equation}
g=\frac{\tau_{+}-2}{2D}+2=\frac{\allowbreak\left(  2\bar{n}+1\right)  }%
{D}\left(  \bar{n}-\sinh^{2}r\right)  , \tag{C2}%
\end{equation}
which leads to
\begin{equation}
g^{2}-\frac{\tau_{-}^{2}}{4D^{2}}=\frac{\allowbreak\left(  2\bar{n}+1\right)
^{2}}{D}. \tag{C3}%
\end{equation}
Then using the integration formula (B7), we can write Eq.(C1) as following
form,%
\begin{align}
W\left(  \alpha,\alpha^{\ast}\right)   &  =\frac{C_{m}^{-1}e^{-2\left\vert
\alpha\right\vert ^{2}}}{\pi\allowbreak\left(  2\bar{n}+1\right)  }%
\frac{\partial^{2m}}{\partial k^{m}\partial t^{m}}\exp\left[  g_{1}\left(
2\alpha+k\right)  \left(  2\alpha^{\ast}+t\right)  \right. \nonumber\\
&  \left.  +\frac{g_{2}}{4}\left(  \left(  2\alpha+k\right)  ^{2}+\left(
2\alpha^{\ast}+t\right)  ^{2}\right)  \right]  _{k=t=0}\nonumber\\
&  =F_{m}\left(  \alpha,\alpha^{\ast}\right)  W_{0}\left(  \alpha,\alpha
^{\ast}\right)  , \tag{C4}%
\end{align}
where $W_{0}\left(  \alpha,\alpha^{\ast}\right)  $ is the WF of squeezed
thermal state defined in Eq.(\ref{f28}), and%
\begin{equation}
\bar{\alpha}=2g_{1}\alpha^{\ast}+g_{2}\alpha,\text{ }g_{1}=\frac{\bar{n}%
-\sinh^{2}r}{2\bar{n}+1},\text{ }g_{2}=\frac{\sinh2r}{2\bar{n}+1}, \tag{C5}%
\end{equation}
as well as%
\begin{equation}
F_{m}\left(  \alpha,\alpha^{\ast}\right)  =C_{m}^{-1}\frac{\partial^{2m}%
}{\partial k^{m}\partial t^{m}}\exp\left[  \bar{\alpha}k+\bar{\alpha}^{\ast
}\allowbreak t+\frac{g_{2}}{4}\left(  k^{2}+t^{2}\right)  +g_{1}kt\right]
_{k=t=0}. \tag{C6}%
\end{equation}
\ Further expanding the exponential term $kt$ included in (C6) into sum
series, and using the generating function of single-variable Hermite polynomials,%

\begin{equation}
H_{n}(x)=\left.  \frac{\partial^{n}}{\partial t^{n}}\exp\left(  2xt-t^{2}%
\right)  \right\vert _{t=0}, \tag{C7}%
\end{equation}
which leads to%
\begin{equation}
\left.  \frac{\partial^{n}}{\partial t^{n}}\exp\left(  At+Bt^{2}\right)
\right\vert _{t=0}=\left(  i\sqrt{B}\right)  ^{n}H_{n}\left[  A/(2i\sqrt
{B})\right]  =\left(  -i\sqrt{B}\right)  ^{n}H_{n}\left[  A/(-2i\sqrt
{B})\right]  , \tag{C8}%
\end{equation}
thus we can see%
\begin{align}
F_{m}\left(  \alpha,\alpha^{\ast}\right)   &  =C_{m}^{-1}\sum_{l=0}^{\infty
}\frac{g_{1}^{l}}{l!}\frac{\partial^{2l}}{\partial\bar{\alpha}^{l}\partial
\bar{\alpha}^{\ast l}}\frac{\partial^{2m}}{\partial k^{m}\partial t^{m}}%
\exp\left[  \bar{\alpha}k+\bar{\alpha}^{\ast}\allowbreak t+\frac{g_{2}}%
{4}\left(  k^{2}+t^{2}\right)  \right]  _{k=t=0}\nonumber\\
&  =\frac{C_{m}^{-1}}{2^{2m}}g_{2}^{m}\sum_{l=0}^{\infty}\frac{g_{1}^{l}}%
{l!}\frac{\partial^{2l}}{\partial\bar{\alpha}^{l}\partial\bar{\alpha}^{\ast
l}}H_{m}(\bar{\beta})H_{m}(\bar{\beta}^{\ast}), \tag{C9}%
\end{align}
where
\begin{equation}
\bar{\beta}=\frac{\sqrt{\allowbreak2\bar{n}+1}}{i\sqrt{\sinh2r}}\bar{\alpha
}=\frac{2\alpha^{\ast}\left(  \bar{n}-\sinh^{2}r\right)  +\alpha\sinh
2r}{i\sqrt{\left(  2\bar{n}+1\right)  \sinh2r}}. \tag{C10}%
\end{equation}
Then using the recurrence relation of $H_{n}(x),$
\begin{equation}
\frac{\mathtt{d}}{\mathtt{d}x^{l}}H_{n}(x)=\frac{2^{l}n!}{\left(  n-l\right)
!}H_{n-l}(x), \tag{C11}%
\end{equation}
Eq.(C9) becomes%
\begin{align}
F_{m}\left(  \alpha,\alpha^{\ast}\right)   &  =\frac{C_{m}^{-1}}{2^{2m}}%
g_{2}^{m}\sum_{l=0}^{\infty}\frac{1}{l!}\left(  \frac{\bar{n}-\sinh^{2}%
r}{\sinh2r}\right)  ^{l}\frac{\partial^{2l}}{\partial\bar{\beta}^{l}%
\partial\bar{\beta}^{\ast l}}H_{m}(\bar{\beta})H_{m}(\bar{\beta}^{\ast
})\nonumber\\
&  =\frac{\left(  m!\right)  ^{2}g_{2}^{m}}{2^{2m}C_{m}}\sum_{l=0}^{m}%
\frac{2^{2l}\left(  \bar{n}-\sinh^{2}r\right)  ^{l}}{l!\left[  \left(
m-l\right)  !\right]  ^{2}\sinh^{l}2r}\left\vert H_{m-l}(\bar{\beta
})\right\vert ^{2}=\text{Eq.}(\ref{f29}). \tag{C12}%
\end{align}
Thus we complete the derivation of WF Eq.(\textbf{\ref{f8}}) by combing Eqs.
(C4) \ and (C12).

\textbf{APPENDIX D:} \textbf{Derivation of (\ref{f35})}

Substituting Eqs.(\ref{f27})-(\ref{f29}) into (\ref{f34}), we have%
\begin{align}
W\left(  \zeta,\zeta^{\ast},t\right)   &  =\frac{2C_{m}^{-1}/\allowbreak
\left(  \allowbreak2\bar{n}+1\right)  }{\pi\left(  2\mathfrak{N}+1\right)
T}\exp\left[  \frac{-2\allowbreak\left\vert \zeta\right\vert ^{2}}{\left(
2\mathfrak{N}+1\right)  T}\right]  \frac{\partial^{2m}}{\partial k^{m}%
\partial\tau^{m}}\exp\left[  g_{1}k\tau+\frac{g_{2}}{4}\left(  k^{2}+\tau
^{2}\right)  \right] \nonumber\\
&  \times\int\frac{d^{2}\alpha}{\pi}\exp\left[  -\left(  2g_{0}+g_{3}%
e^{-\kappa t}\right)  \left\vert \alpha\right\vert ^{2}+\left(  2\allowbreak
\tau g_{1}+kg_{2}+g_{3}\zeta^{\ast}\right)  \alpha\right. \nonumber\\
&  +\left.  \left(  2kg_{1}+\allowbreak\tau g_{2}+g_{3}\zeta\right)
\alpha^{\ast}+g_{2}\left(  \alpha^{2}+\alpha^{\ast}{}^{2}\right)  \right]
_{k=\tau=0}\nonumber\\
&  =\frac{C_{m}^{-1}/\allowbreak\left(  \allowbreak2\bar{n}+1\right)  }%
{\pi\left(  2\mathfrak{N}+1\right)  T\sqrt{G}}\exp\left[  -\Delta
_{2}\left\vert \zeta\right\vert ^{2}+\frac{\allowbreak g_{2}g_{3}^{2}}%
{4G}\left(  \zeta^{2}+\zeta^{\ast2}\right)  \right] \nonumber\\
&  \times\frac{\partial^{2m}}{\partial k^{m}\partial\tau^{m}}\exp\left[  \chi
k\tau+\omega\allowbreak^{\ast}k+\omega\tau+\Delta_{1}\left(  k^{2}+\tau
^{2}\right)  \right]  _{k=\tau=0}, \tag{D1}%
\end{align}
where $T=(1-e^{-2\kappa t},$ ($g_{0},g_{1},g_{2}$, $g_{3})$ and ($\chi
,\omega,G,\Delta_{1},\Delta_{2})$ are defined in Eqs.(\ref{f38}) and
(\ref{f39}), respectively. In a similar way to deriving Eq.\textbf{(\ref{f8}),
}we can further put Eq.(D1) into Eqs.(\ref{f35})-(\ref{f37}).

\textbf{APPENDIX\ E: Derivation of (\ref{f47})}

Then employing Eqs.(\ref{f28}) and (C6) as well as the integration formula
(B7), we can treat the integration in a similar way to deriving Eq.(\ref{f8}),%
\begin{align}
\mathtt{tr}\left(  \rho_{s}\rho\right)   &  =\frac{4C_{m}^{-1}}{\allowbreak
\left(  \allowbreak2\bar{n}+1\right)  \allowbreak^{2}}\frac{\partial^{2m}%
}{\partial k^{m}\partial t^{m}}\exp\left[  \frac{g_{2}}{4}\left(  k^{2}%
+t^{2}\right)  +g_{1}kt\right] \nonumber\\
&  \times\int\frac{d^{2}\alpha}{\pi}\exp\left[  -4g_{0}\left\vert
\alpha\right\vert ^{2}+\left(  kg_{2}+2tg_{1}\right)  \alpha+\left(
2kg_{1}+tg_{2}\right)  \alpha^{\ast}+2g_{2}\left(  \alpha^{2}+\alpha^{\ast}%
{}^{2}\right)  \right]  _{k=t=0}\nonumber\\
&  =\frac{C_{m}^{-1}}{\allowbreak\left(  \allowbreak2\bar{n}+1\right)
\allowbreak^{2}\sqrt{g_{0}^{2}-g_{2}^{2}}}\frac{\partial^{2m}}{\partial
k^{m}\partial t^{m}}\exp\left[  \frac{g_{2}}{4}\left(  k^{2}+t^{2}\right)
+g_{1}kt\right] \nonumber\\
&  \times\exp\left.  \left[  \frac{g_{2}\left(  4g_{1}^{2}+4g_{0}g_{1}%
+g_{2}^{2}\right)  \allowbreak}{8\left(  g_{0}^{2}-g_{2}^{2}\right)  }\left(
k^{2}+t^{2}\right)  +\frac{4g_{0}g_{1}^{2}+4g_{1}g_{2}^{2}+g_{0}g_{2}^{2}%
}{4\left(  g_{0}^{2}-g_{2}^{2}\right)  }kt\right]  \right\vert _{k=t=0}%
\nonumber\\
&  =\frac{C_{m}^{-1}}{\allowbreak\left(  \allowbreak2\bar{n}+1\right)
\allowbreak^{2}\sqrt{g_{0}^{2}-g_{2}^{2}}}\left.  \frac{\partial^{2m}%
}{\partial k^{m}\partial t^{m}}\exp\left[  B_{2}^{\prime}\left(  k^{2}%
+t^{2}\right)  +\allowbreak B_{1}^{\prime}kt\right]  \right\vert _{k=t=0},
\tag{E1}%
\end{align}
where $g_{0}^{2}-g_{2}^{2}=\frac{1}{\left(  2\bar{n}+1\right)  ^{2}}$ and
\begin{align}
B_{1}  &  =\frac{1}{4}\frac{g_{0}}{g_{0}^{2}-g_{2}^{2}}\left(  4g_{1}%
^{2}+4g_{0}g_{1}+g_{2}^{2}\right) \nonumber\\
&  =\frac{\allowbreak\bar{n}\left(  \bar{n}+1\right)  }{2\bar{n}+1}%
\cosh2r=g_{0}\allowbreak\bar{n}\left(  \bar{n}+1\right)  ,\tag{E2}\\
B_{2}^{\prime}  &  =\frac{1}{8}\frac{g_{2}\left(  2g_{0}^{2}+4g_{0}%
g_{1}+4g_{1}^{2}-g_{2}^{2}\right)  }{g_{0}^{2}-g_{2}^{2}}\nonumber\\
&  =\allowbreak\frac{2\bar{n}^{2}+2\bar{n}+1}{4\left(  2\bar{n}+1\right)
}\sinh2r=\frac{g_{2}}{4}\left(  2\bar{n}^{2}+2\bar{n}+1\right)  . \tag{E3}%
\end{align}
Similarly to deriving Eq.(B11), we have%
\begin{equation}
\left.  \frac{\partial^{2m}}{\partial k^{m}\partial t^{m}}\exp\left[
B_{2}^{\prime}\left(  k^{2}+t^{2}\right)  +\allowbreak B_{1}kt\right]
\right\vert _{k=t=0}=m!B_{2}^{m/2}P_{m}\left(  B_{1}/\sqrt{B_{2}}\right)  ,
\tag{E4}%
\end{equation}
and $B_{2}\equiv B_{1}^{2}-4B_{2}^{\prime2}$ given in Eq.(\ref{f48}), which
leads to Eq.(\ref{f47}).

\end{document}